\documentclass[aps,prx,twocolumn,floatfix,superscriptaddress]{revtex4-2}

\usepackage{amsmath}  
\usepackage{amsfonts} 
\usepackage{graphicx} 
\usepackage{epstopdf}
\usepackage{gensymb}
\usepackage{color}
\usepackage{soul}

\begin{document}

\title{Acoustically levitated lock and key grains} 
\author{Melody X. Lim}
\email{mxlim@uchicago.edu}
\affiliation{James Franck Institute, The University of Chicago, Chicago, Illinois 60637, USA}
\affiliation{Department of Physics, The University of Chicago, Chicago, Illinois 60637, USA}
\author{Heinrich M. Jaeger}
\affiliation{James Franck Institute, The University of Chicago, Chicago, Illinois 60637, USA}
\affiliation{Department of Physics, The University of Chicago, Chicago, Illinois 60637, USA}

\begin{abstract}
We present a scheme for generating shape-dependent, specific bonds between millimeter scale particles, using acoustic levitation. We levitate particles in an ultrasonic standing wave, allowing for substrate-free assembly. Secondary scattering generates shape-dependent attractive forces between particles, while driving the acoustic trap above its resonance frequency produces active fluctuations that mimic an effective temperature. We 3D print planar particles, and show that the local curvature of their binding sites controls the selectivity for attaching a matching particle. We find that the bound-state probability and bound-state lifetime can be independently tuned via the binding site depth and height respectively. Finally, we show that these principles can be used to design particles that assemble into complex structures.  
\end{abstract}
\maketitle

The self-assembly of materials with complex structure requires mechanisms for programmable, specific bonds between subunits. Such directed interactions have been successfully developed in systems of self-assembling nanoparticles and colloidal particles, relying on the addressability of hydrogen-bonded complementary DNA strands~\cite{wang2012colloids,wang2015crystallization,mcmullen2018freely}, or the shape~\cite{sacanna2010lock,jones2010dna,van2014understanding,lu2015superlattices} and surface~\cite{kraft2012surface,oh2019colloidal} dependence of depletion (entropic)  and electrostatic~\cite{demirors2015long, mihut2017assembling} forces. However, there remains a need for equivalent strategies for directed bonds, with tunable specificity, between out-of-equilibrium particles. Prior approaches to directed self-assembly between macroscale particles have made use of magnetically patterned particles~\cite{niu2019magnetic,gu2019magnetic}, capillary forces~\cite{wang2017dynamic}, and molecular recognition between surface-functionalized hydrogel particles~\cite{harada2011macroscopic}. Here, we show that acoustic levitation offers one possibility for the generation of material-agnostic, substrate-free tunable forces between granular particles~\cite{lim2019cluster,lim2022mechanical}. 

Previous work has shown that the local curvature of a granular particle controls the strength of interparticle acoustic forces, resulting in flexible hinge-like structures~\cite{lim2019edges}. We extend this work to show that the in-plane shape of a planar particle can also be used to tune local binding probabilities between a ``lock" particle and a matching ``key", via the shape dependence of the secondary acoustic scattering forces. We take advantage of 3D printing to fabricate and rapidly prototype precisely designed particles. Our results demonstrate that the local curvature of a particle binding site controls the selectivity for attaching a matching particle. The bound-state lifetime can further be tuned by adjusting the height of the particles. We show that these principles can be used to design particles that assemble into complex structures.  

\begin{figure}
\includegraphics[width = 1\columnwidth]{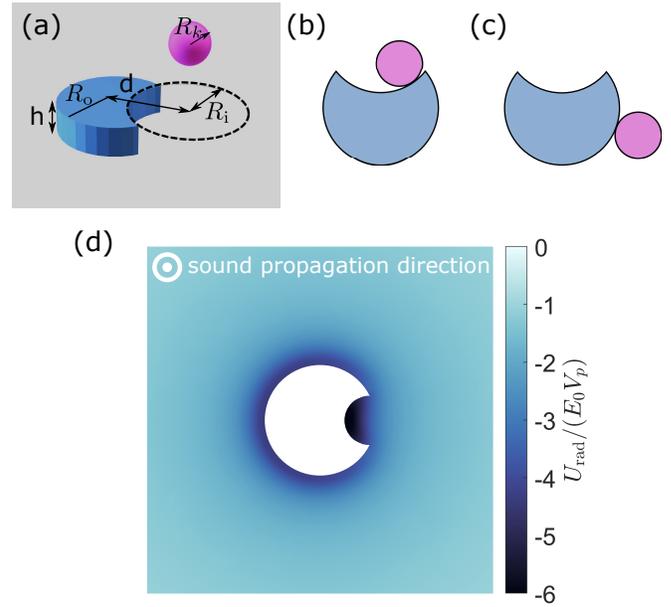}
\caption
{\textbf{Schematic of the lock-key system.} (a) 3D drawing of a lock (blue) and key (magenta) particle, labelled with lock shape parameters. For all clusters presented here,~$R_o = 1$mm, and~$R_k = 0.355-0.425$mm. We explore the effects of $d$, $R_i$ and $h$ on the probability that the key particle binds to the lock binding site. Successful ``eating" is defined by the key particle being attached anywhere on the lock interior (b). Alternatively, a key particle can be attached to the lock exterior (c). (d) Finite-element simulations of the acoustic potential~$U_\mathrm{rad}$ in the nodal plane (lock particle shown in white, with~$R_i=0.5$mm, $h = 0.5$mm, and $d/R_i=1$), normalized by the acoustic energy per key particle volume~$E_0 V_k$. }
\label{fig:cresc_coords}
\end{figure}

Our granular locks are crescent shaped, composed of the union of a disk and a disk-shaped hole, whose centers are separated by distance~$d$ (see Fig.~\ref{fig:cresc_coords}(a) for a schematic). We fix the radius of the disk to be $R_o = 1$mm, with the radius of the disk-shaped hole $R_i$, and the height~$h$ of the crescent variable. The locks were 3D printed using UV-cured hard plastic (Vero family of materials, density 1170-1180 kg m$^{-3}$) using a Stratasys J850 Prime printer (resolution $\pm 100 \mu$m).  For keys we use polyethylene spheres (Cospheric, density 1000 kg m$^{-3}$, diameter $2R_k=710–850\mu$m). 

All particles are levitated in an acoustic trap in air, with sound wavelength~$\lambda_0=7.5$mm and trap height~$\lambda_0$. Lock particles levitate in the lower of the two nodal planes with their flat faces roughly parallel to the reflector and transducer, $\lambda_0/4$ above the reflector surface. When a key particle is inserted with a pair of tweezers, the secondary scattering between the particles generates secondary acoustic forces, which bind the lock and key as a compact cluster. 

For particles much smaller than the wavelength of sound, these secondary acoustic forces can be expressed as the gradient of an acoustic potential, which is derived from scattering expansions of the pressure and velocity fields in the acoustic cavity~\cite{bruus2012acoustofluidics, silva2014acoustic}. The pressure and velocity fields resulting from a given set of boundary conditions (including the presence of another particle) can then be computed using finite element simulations (COMSOL Multiphysics here), and combined to give the acoustic potential around the particle perimeter,~$U_\mathrm{rad}$. In this case, we fix a perfectly scattering lock particle in the levitation plane, and set up a standing wave (between perfectly reflecting transducer and reflector) whose frequency and wavelength match the experiment.  This standing wave produces a cavity with acoustic energy density~$E_0$, which combined with the key particle volume~$V_k$ produces a characteristic energy scale. Acoustic fields are dissipated by plane-wave radiation conditions on the lateral boundaries. An example of the acoustic radiation potential in the nodal plane around a lock particle is shown in Fig.~\ref{fig:cresc_coords}(d). The perimeter of the lock particle is characterized by a relatively uniform binding energy, which rapidly decays away from the surface of the lock particle. 

In order to allow the key particle to explore the lock particle perimeter, we drive the acoustic trap with a frequency larger than its resonance frequency ($\Delta f/f_0=2.5\times 10^{-3}$). When small clusters of spherical particles are levitated in such a far-from-resonance trap, they rearrange by moving particles randomly along their periphery~\cite{lim2019cluster}, producing cluster configuration statistics that are indistinguishable from the statistics of clusters of colloidal particles in thermal equilibrium. Here, we use this far-from-resonant acoustic trap to produce an effective temperature in clusters of lock-key particles, anticipating that the statistics produced can also be understood via parallels to clusters of thermal colloids. 

\begin{figure*}
\includegraphics[width = 2\columnwidth]{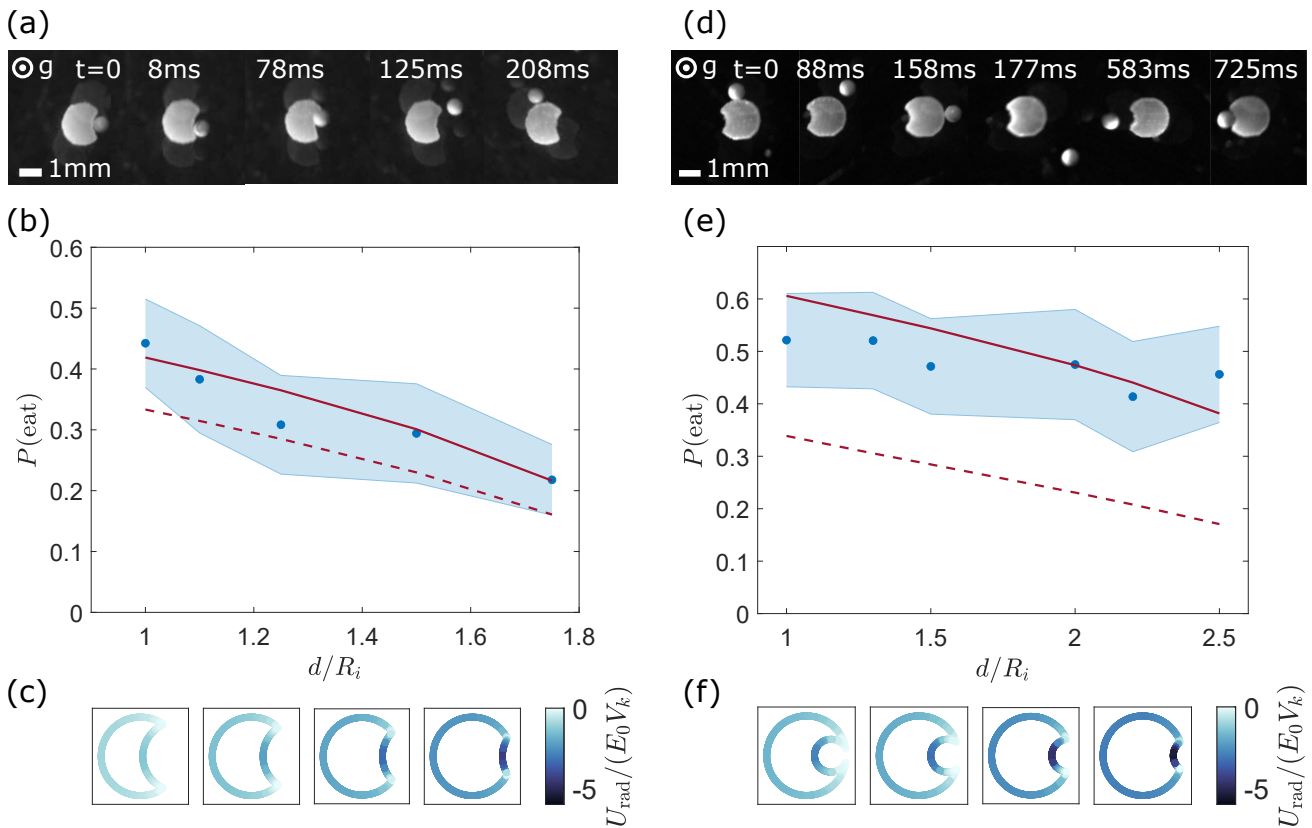}
\caption{
\textbf{The match between the lock binding site curvature and the key controls preferential key attachment.}  For data in this figure,~$h=0.5$mm. (a-c): When the lock binding site curvature does not match the key, key statistics are consistent with an unweighted random walk around the lock perimeter. (a) Time-series of images showing the escape of a key particle from a lock binding site with~$d/R_i=1.5$, $R_i=1$mm,~$h=0.5$mm. (b) Probability of ``eating" a key particle for a lock particle whose binding site is twice the radius of a key particle, as a function of the overlap distance~$d$ divided by the binding site radius~$R_i=1$mm (data plotted in blue, with shaded standard error). Maroon solid (dashed) line indicates the result of Eq.~\ref{eq:Pg_unweighted} with (without) a geometric correction to~$R_i$ and~$R_o$, see text for details. (c) Finite-element simulations of the acoustic potential~$U_\mathrm{rad}$, normalized by the acoustic energy per key particle volume~$E_0 V_k$ (from left to right)~$d/R_i= 1, \: 1.25,\:  1.5,\:  1.75$.  (d--f): When the lock binding site curvature matches the key, the probability of successful binding is increased. (d) Time-series of images showing the entrance of a key particle to a lock binding site with~$d/R_i=1.5$, $R_i=0.5$mm. (e) Probability of ``eating" a key particle for a lock particle whose binding site is the same radius as a key particle, as a function of the overlap distance~$d$ divided by the binding site radius~$R_i$ (data plotted in blue, with standard error indicated by the shaded regions). Maroon solid (dashed) line indicates the result of Eq.~\ref{eq:Pg_unweighted} with (without) a geometric correction to~$R_i$ and~$R_o$, see text for details. (f) Finite-element simulations of the acoustic potential on the particle perimeter for (from left to right)~$d/R_i= 1, \: 1.5,\:  2,\:  2.5$. The colorscale is shared between all four plots (and also part (c) of this figure), and shows the acoustic potential on the lock edge~$U_\mathrm{rad}$, normalized by the acoustic energy per key particle volume~$E_0 V_k$.}
\label{fig:cresc}
\end{figure*}

As a result of these acoustic fluctuations, the key particle shuffles around the rim of the lock particle, occasionally detaching from the lock and orbiting, before bouncing against the lock and again becoming bound to the lock particle (see Fig.~\ref{fig:cresc}(a) and Fig.~\ref{fig:cresc}(d) for images of the cluster rearrangement dynamics, with corresponding movies in Supplementary Movies 1 and 2). To record the relative in-plane positions of the lock and key particle, a bottom view of the cluster dynamics was recorded with a high-speed camera (Vision Research Phantom v25) at 3,000 frames per second. Recordings stopped when an excessively energetic rearrangement ejected the lock or key particle from the acoustic trap. To average across variation in the production process, rearrangements were recorded with thirty different lock and key pairs from the same 3D-printed batch. 

 During these cluster rearrangement events, the key particle explores the outer perimeter of the lock, including the cut-out binding site. We assess the likelihood that the lock particle successfully captures (``eats") the key particle in its binding site by measuring the probability of eating~$P(\mathrm{eat})$, defined as the ratio of the time spent with the key attached to the inside of the binding site (see Fig.~\ref{fig:cresc_coords}(b) and (c) for definitions of inside and outside), to the total time with the key attached to the lock at any location. 

When the lock binding site is roughly twice the radius of the key ($R_i=1$mm, images shown in Fig.~\ref{fig:cresc}(a)), we find that~$P(\mathrm{eat})$ decreases with the area of the binding site. Specifically, as~$d/R_i$ increases,~$P(\mathrm{eat})$ decreases, as does the fraction of the particle perimeter associated with the binding site (Fig.~\ref{fig:cresc}(b)). To gauge the extent to which this decrease in the binding site area contributes to the observed binding probabilities, we compute a geometric estimate for the probability of eating,~$P_g$. We assume that the key particle is equally likely to bind to any position on the particle perimeter, such that the probability of binding is the fraction of the total lock perimeter occupied by the binding site. Quantitatively, 

\begin{align}
    P_g &= \frac{R_i \theta_b}{R_o (2\pi - \theta_b)+ R_i \theta_b}    
    \label{eq:Pg_unweighted}
    \end{align}
    where
    
\begin{align*}
    \theta_b &= 2\arccos\left[  \frac{R_o^2 +d^2 -R_i^2}{2 d R_o}\right] \:.
\end{align*}
Comparing the experimental data to~$P_g$ (maroon dotted line in Fig.~\ref{fig:cresc}(b)) reveals that~$P_g$ is compatible with much of the experimental data, with the exception of the smallest~$d/R_i$. For the case where the lock and key do not match, then our observed experimental binding statistics are consistent with an unbiased random walk of the key around the lock particle perimeter. 

In order to introduce preferential binding, we turn to the radius of the lock binding site. Matching the radius of the key particle by shrinking $R_i$ to 0.5mm results in~$P(\mathrm{eat}) \approx 0.5$ (images in Fig.~\ref{fig:cresc}(d)). For matching keys and locks, we observe values for~$P(\mathrm{eat})$ that are more than double that of the geometric expectation~$P_g$ (Fig.~\ref{fig:cresc}(e)). These results strongly suggest that the match in radius of curvature between the lock and key particles produces preferential binding at the lock site. 

\begin{figure}
\includegraphics[width = 0.9\columnwidth]{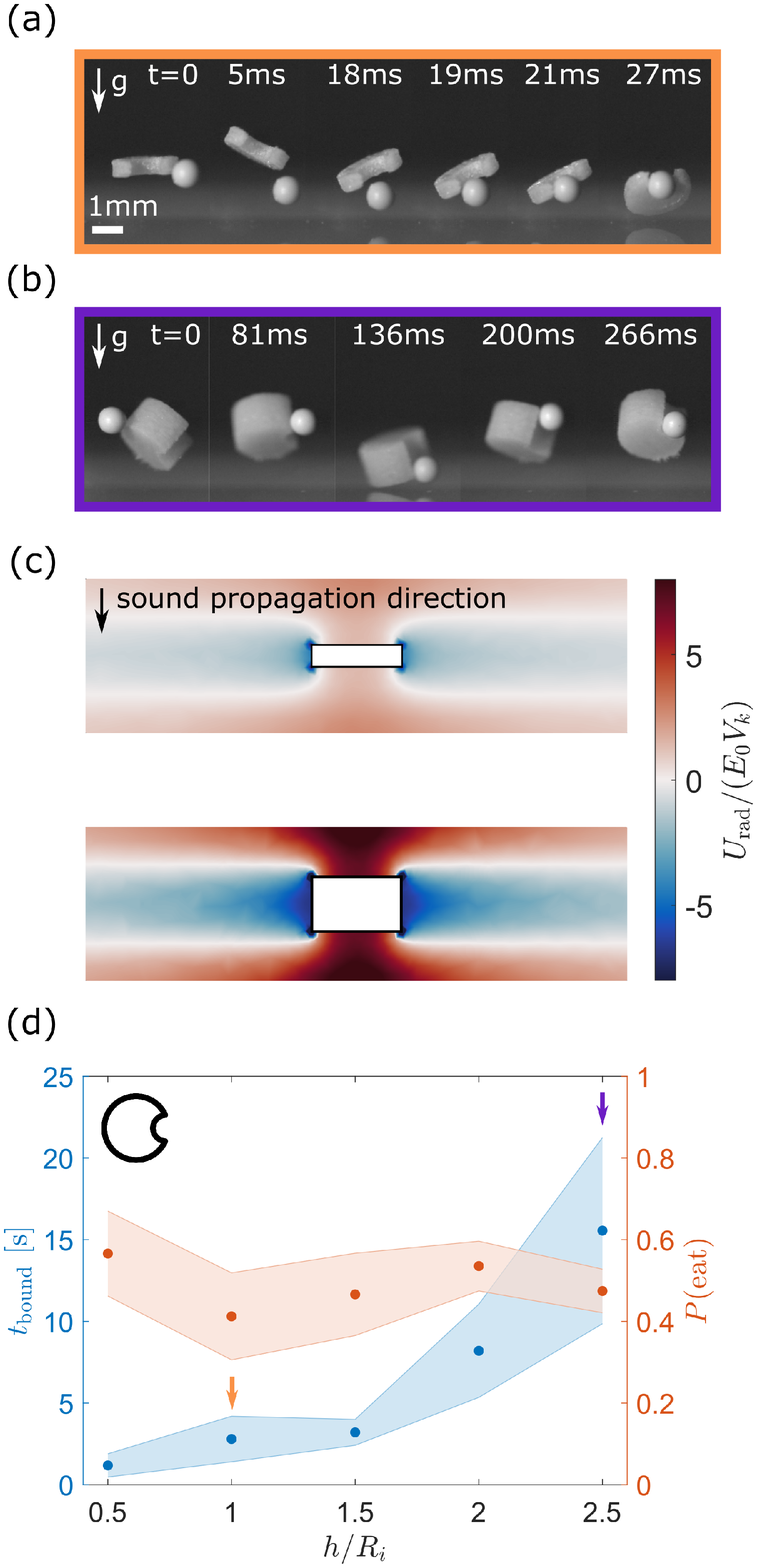}
\caption{\textbf{Lock particle height tunes the bound-state lifetime, without tuning the bound state probability.} (a) Time-series of side-view images, showing a key particle entering the binding site of a lock with~$d/R_i=1$, $R_i=$0.5mm, $h=0.5$mm. (b) Time-series of side-view images, showing a key particle entering the binding site of a lock with~$d/R_i=2$, $R_i=$0.5mm, $h=1.25$mm. (c) Normalized acoustic potential energy around a vertical cross-section of a lock particle with (top) $h=$0.5mm, and (bottom) $h=$2.5mm
(d) Probability of ``eating" a key particle for a lock particle with~$d/R_i=2$, $R_i=$1mm (orange data points, right axis) and average bound state lifetime (blue data points, left axis), as a function of the key height~$h$ divided by the binding site radius~$R_i$. A schematic of the projected lock geometry is shown in black. Shaded areas indicate the standard error. Data points corresponding to the still images in (a) and (b) are marked with colored arrows. }
\label{fig:height}
\end{figure}

One possibility is that the matching curvature of the lock and key increases the lock binding site energy, since acoustic scattering is strongly shape dependent. In order to measure this effect, we simulate the energy landscape of the lock perimeter. The results of these finite-element simulations are shown in Figs.~\ref{fig:cresc}(c) and (f), for both the matched and unmatched lock-key pair. The energetic landscape indeed varies strongly around the perimeter of each lock particle, with local minimums induced at the center of the binding sites, and strong local maxima at the sharp tips of the binding site.  However, the change in binding site curvature does not increase the depth of the energetic minimum at the binding site center (compare Fig.~\ref{fig:cresc}(c) and (f)). The discrepancy in~$P(\mathrm{eat})$ therefore cannot be accounted for by changes in the acoustic binding energy of the lock-key cluster.   

In order to account for this discrepancy, we instead turn to the geometry of the contact between the lock and key particles. Eq.~\ref{eq:Pg_unweighted} assumes that all arc-lengths should be treated with equal weighting between the inside and outside of the lock. In the case where the key particle is small compared to the binding site, this is an appropriate assumption, especially as the acoustic force is short-ranged (most of the acoustic binding energy is contained within a radius of the key particle, as shown in Fig.~\ref{fig:cresc_coords}(d)). However, the area of contact between a key and a matching binding site is much larger, meaning that the effective length of the binding site increases as the match in key and lock becomes closer. We derive this effective radius by considering the distance between the lock and key surfaces at a given point of contact~\cite{white1983deryaguin,israelachvili2011intermolecular, sacanna2010lock}. For the lock inside (which has radii of curvature 0 and~$-R_i$), touching a key (for which both radii of curvature are~$R_k$), this effective radius is  

\begin{align}
    R_\mathrm{eff,i} &= \sqrt{\frac{R_k^2 R_i}{R_i - R_k}} \: .
    \label{eq:Reffin}
\end{align}
Similarly, the lock outside has radii of curvature 0 and~$R_o$, producing the following effective radius: 

\begin{align}
    R_\mathrm{eff,o} &= \sqrt{\frac{R_k^2 R_o}{R_o + R_k}} \: .
    \label{eq:Reffout}
\end{align}

Using $R_\mathrm{eff,o}$ and $R_\mathrm{eff,i}$ instead of~$R_i$ and~$R_o$ in  Eq.~\ref{eq:Pg_unweighted} produces the maroon solid lines in Fig.~\ref{fig:cresc}(b) and (e). These predictions are compatible with the experimentally observed binding probabilities for both shapes over the range of observed~$d/R_i$, and reproduce the difference between the matched and unmatched binding probabilities. Our results thus suggest that the probability of binding is controlled by the match in curvature between the lock and binding site, as parameterized by Eq.~\ref{eq:Reffin} and Eq.~\ref{eq:Reffout}. This geometric correction to the binding probabilities is reminiscent of the use of the Derjaguin approximation in calculations of the depletion force in colloidal systems~\cite{sacanna2010lock}, raising the possibility of further analogies between acoustic forces and other types of thermal fluctuation forces.

\begin{figure*}
\includegraphics[width = 2\columnwidth]{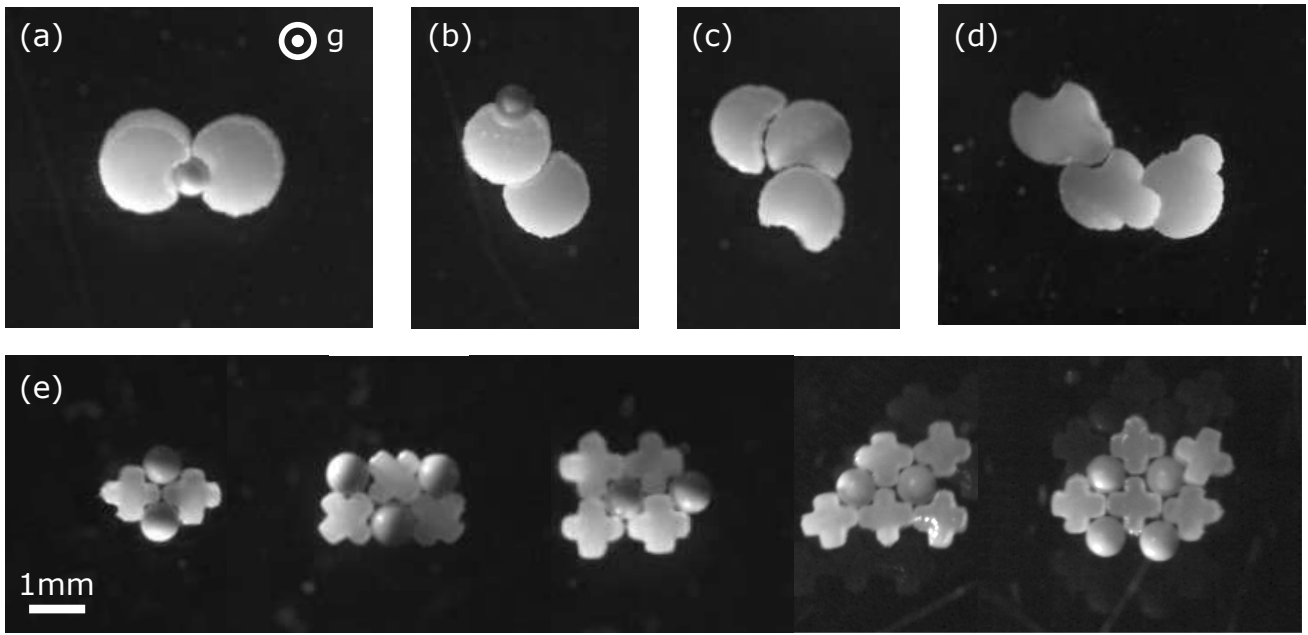} 
\caption
{\textbf{Self-assembly of lock and key granular clusters.} (a) Cluster composed of two lock particles and a key. (b) Three lock particles with additional keys printed on the outside. (c) A key, with two locks of varying binding site curvature. (d) Three keys with varying binding site overlap distance. (e) Spherical particles with cross-shaped particles form alternating, ``host-guest" structures with (i) two spheres and two crosses, (ii) three spheres and three crosses, (iii) two spheres and four crosses, (iv) two spheres and five crosses, and (v) four spheres and five crosses. (b) A spherical particle with a pair of lock particles. }
\label{fig:structures}
\end{figure*}

Although the statistics of the binding site occupancy are directly controlled by the in-plane geometry of the lock perimeter, the full cluster dynamics take place in a three-dimensional acoustic trap, where particles are weakly confined to lie in the levitation plane. A particularly interesting case is presented by~$d/R_i=1$, when $R_i<R_o$. In this case, where the binding site is completely enclosed in the lock, a random walk around the outside of the lock particle would not be expected to result in a successful binding between the lock and key particle. The fact that these binding events are observed with high probability suggests that the transition between bound and unbound keys must involve some degree of motion in three dimensions: i.e. that the cluster does not remain fully parallel to the levitation plane as it fluctuates. Such a transition between bound and unbound states involves a transition of the key over the lock edge, and onto the face of the lock. An example of such an event is shown in Fig.~\ref{fig:height}(a) (see Supplementary Movie 3 for dynamics), where a key enters the binding site by going beneath the tips of the lock particle.

These three-dimensional crossing events imply another possibility for tuning the bound-state lifetime of a lock and key cluster: the height of the lock particle. In addition to particle flat faces being repulsive~\cite{lim2019edges}, the out-of-plane secondary acoustic forces that keep clusters planar scale with the particle volume (Fig.~\ref{fig:height}(c)). Finally, particles that escape out of a tall binding site face the additional energetic cost of the primary acoustic field, which acts to keep the particle in the nodal plane. In addition, inside a binding site larger than the diameter of the key particle, a key can fluctuate between the two vertical edges of the binding site (Fig.~\ref{fig:height}(b), see Supplementary Movie 4 for dynamics), offering both a means to dissipate the kinetic energy injected by the active fluctuations, and also a reduction in the probability of a direct collision between the key particle and the reflector plate.  

These effects lead to an increase in the bound state lifetime for a lock-key pair by up to a factor of four (statistics and average bound time in Fig.~\ref{fig:height}(d)). Unlike colloidal clusters in thermal equilibrium, where increasing the bound state lifetime by decreasing the temperature leads to a corresponding increase in bound state probability, increasing the bound state lifetime by tuning the lock thickness does not change the probability of the key being eaten by the lock. The three-dimensional nature of the acoustic lock and key cluster thus offers an opportunity to separately tune the bound state probabilities and lifetimes, to yield designed structures with parts whose dynamics can take place over a wide range of timescales. 

Using these design principles, we can take advantage of the flexible acoustic bonds between lock and key to create a variety of lock and key granular clusters (Fig.~\ref{fig:structures}). A pair of locks can assemble on either side of a key, forming a flexible jointed structure in the levitation plane (Fig.~\ref{fig:structures}(a)). Alternatively, a lock particle can attach to the outside of a lock particle with matching radius of curvature. Combined with a key, the cluster of three particles can form a snowman-like stack of disks (Fig.~\ref{fig:structures}(b)). The match between the curvature of the lock outside and a different lock binding site can also result in stacked chains of particles in the absence of keys (Fig.~\ref{fig:structures}(c)). 

Additional features can also be easily added to the lock. For instance, the position of the key can be directly fixed onto the lock body by printing an additional intersecting disk, again resulting in chains of particles (Fig.~\ref{fig:structures}(d)). Varying the position of the attached key can then be used to tune the persistence length of the granular chain. Alternatively, additional binding sites can be added to the lock surface. These can then be combined with additional keys and locks to form host-guest clusters (Fig~\ref{fig:structures}(e), see Supplementary Movie 5 for dynamics). 

Here, we have taken the first steps towards acoustically levitated directed assembly. We have shown that shape plays a key role in regulating the assembly probabilities of lock and key structures: when the radius of curvature of a key and lock binding site match, their probability of binding can increase by a factor of two (compared to the unmatched case). This increased probability of binding was a direct result of the short-ranged acoustic scattering force. Our results show that matching curvatures between acoustically levitated objects directly controls the selectivity of an acoustic bond. In general, such curvature-matching effects can be used to generate selective binding in clusters bound by forces which decay on lengthscales of the order of the particle size.

The effect of lock shape (as parameterized here by binding site depth, binding site curvature, and lock height) on the lock energy landscape can be easily simulated using finite element simulations, then rapidly prototyped using 3D printing. Additional forces can also be incorporated into these millimeter-size grains, for instance by patterning the grain edges with magnetic patches~\cite{snezhko2011magnetic, niu2019magnetic}, inducing acoustohydrodynamic interactions~\cite{abdelaziz2021ultrasonic}, or by levitating the grains in an atmosphere saturated with water vapor to encourage the formation of capillary bridges~\cite{lewandowski2010orientation,bharti2016capillary}. Such an approach is further amenable to algorithmic optimization~\cite{hormoz2011design,pandey2011algorithmic,miskin2013adapting} to produce a desired energy landscape and binding time, opening the door to the systematic exploration and creation of specific, functional granular acoustic interactions. 

\section*{Acknowledgments}
We thank B. VanSaders and J. Kim for useful discussions. We thank T. Gandhi for useful discussions, and help fabricating the particles. 
This research was supported by the National Science Foundation through Grant No. DMR-2104733.
The experiments utilized the shared experimental facilities at the University of Chicago MRSEC, which is funded by the National Science Foundation under award number DMR-2011854. 
This research utilized computational resources and services supported by the Research Computing Center at the University of Chicago.

\subsection*{Appendix: Description of Supplementary Movies}
\textbf{Supplementary Movie 1.} Movie showing the rearrangements of a lock-key pair from below, where the lock binding site curvature does not match the key (see Fig. 2(a) of the main text for corresponding still images). 

\textbf{Supplementary Movie 2.} Movie showing the rearrangements of a lock-key pair from below, where the lock binding site curvature matches the key (see Fig 2(d) of the main text for corresponding still images). 

\textbf{Supplementary Movie 3.} Movies showing the rearrangements of a lock-key pair from the side, where the lock binding site curvature matches the key, and the binding site is partially enclosed within the lock (see Fig. 3(a) for corresponding still images). 

\textbf{Supplementary Movie 4.} Movies showing the rearrangements of a lock-key pair from the side, where the lock height is much larger than the key diameter (see Fig. 3(b) for corresponding still images).

\textbf{Supplementary Movie 5.} Short movie showing the rearrangements of a host-guest structure from below, composed of three crosses and three spheres (corresponding still images in Fig 4(e) of the main text).

\newpage

\end{document}